# Rethinking Eye-blink: Assessing Task Difficulty through Physiological Representation of Spontaneous Blinking

YOUNGJUN CHO*

UCL Interaction Centre, Department of Computer Science, University College London, London, United Kingdom

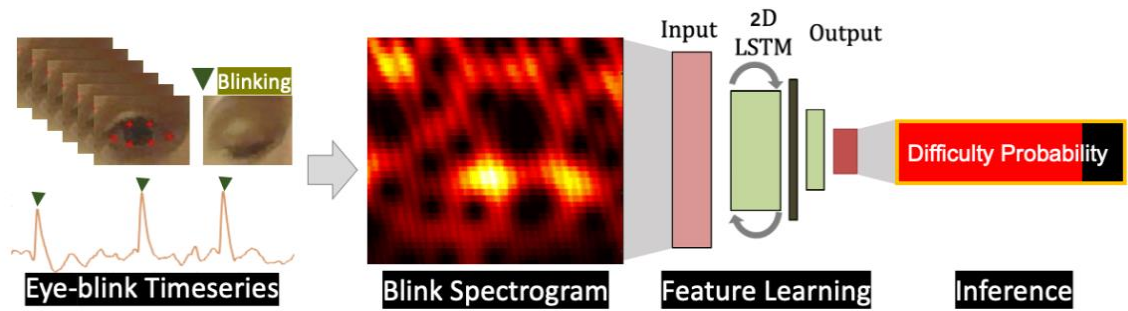

Figure 1: *Rethinking Eye-Blink* is a novel strategy to monitor task difficulty and mental workload continuously and automatically through a physiological representation of a user's spontaneous blinking. This approach uses a standard RGB camera to extract eye-blink timeseries which we represent on time-frequency spectrograms for automatic feature learning.

Continuous assessment of task difficulty and mental workload is essential in improving the usability and accessibility of interactive systems. Eye tracking data has often been investigated to achieve this ability, with reports on the limited role of standard blink metrics. Here, we propose a new approach to the analysis of eye-blink responses for automated estimation of task difficulty. The core module is a time-frequency representation of eye-blink, which aims to capture the richness of information reflected on blinking. In our first study, we show that this method significantly improves the sensitivity to task difficulty. We then demonstrate how to form a framework where the represented patterns are analyzed with multi-dimensional Long Short-Term Memory recurrent neural networks for their non-linear mapping onto difficulty-related parameters. This framework outperformed other methods that used hand-engineered features. This approach works with any built-in camera, without requiring specialized devices. We conclude by discussing how *Rethinking Eye-blink* can benefit real-world applications.

---

* *youngjun.cho@ucl.ac.uk*

## 1 INTRODUCTION

It is of critical importance to assess the task difficulty and mental workload that users experience in improving usability problems of human-computer interaction applications [8,34]. Proper assessment can assist in identifying hidden barriers to the use of computing systems and user interfaces. For instance, virtual reality applications introducing new interactive features may include complicated interfaces. This often increases a user's mental workload significantly, leading to poor usability [24,30]. Moreover, high mental workload appears to be associated with accessibility problems [32]. Carefully monitoring users' mental barriers can be a key to building more accessible user interfaces.

Self-report questionnaires, such as NASA-TLX [35], as well as ethnographic and observational methods [42,55], have helped monitor users' workload and address usability problems in many studies. However, we often face some limitations related to the methods when *scaling up* our data collection with a number of users and contexts where it is difficult to have human observers. With recent advances in sensing interfaces and machine learning, physiological computing can possibly help to continuously assess mental barriers without the need for the presence of data collectors, thus compensating for such limitations. This builds on the fact that physiology reflects human minds and mental states [23,36]. Mental demand triggers the sympathetic nervous system to change our metabolic patterns, e.g. accelerating cardiac pulse and respiration, as well as activating sweat glands. This is followed by the body's reaction to the changed metabolism so as to regulate it [23]. A variety of physiological signatures such as eye movement, heart rate variability and skin conductance response from multiple sensors have been investigated to assess their relationship with a user's mental overload and other states and to even estimate them [36,38,46]. To deploy such capability in real-world settings, it is essential to minimize sensing overhead, such as the number of sensors and their size, to the point where any users can easily use them while maintaining reasonable assessment and inference performance [16,27,61].

The present paper contributes to this body of work by focusing primarily on spontaneous eye-blinking [4] which can be monitored simply through a built-in sensing channel (e.g. a webcam). While eye tracking has been exploited in many interactive systems and physiological computing (e.g. [13,27,47,74]), existing eye-blink measures appear to play a minor role in mental workload and task difficulty assessment [25,58,60]. This can be partially explained by the simplicity of the standard metrics, spontaneous Blink Rate (BR) and Blink Duration (BD), which are not designed to capture complex dynamic patterns in physiology (i.e. variability) [4]. To better capture the richness of information reflected on blinking, we propose i) a time-frequency representation of blinking to condense informative physiological patterns associated with task difficulty in an efficient way, and ii) an automated monitoring system where an advanced machine learning algorithm maps represented patterns onto task difficulty levels. This approach, overviewed in Figure 1, requires no additional device but only a camera on a user's computing device (e.g. laptop, smartphone and head-mounted display). This is critical in continuously assessing the usability problems of many applications deployed in the real world.

**Contributions -** this paper makes three main contributions
- *Rethinking Eye-blink*: a novel framework that focuses on a physiological representation of spontaneous eye-blinking for automated monitoring of mental workload associated with task difficulty in a contact-free manner.

- Report on studies that demonstrate the stronger sensitivity of the blink representation to task difficulty than standard metrics and show that the overall framework outperforms other methods that use hand-engineered features in inference tasks.
- Discussion on potential interactive applications of the Rethinking Eye-blink framework.

## 2 RELATED WORK

In this section we review prior work on interactive systems using eye tracking and physiological computing for mental state monitoring. We then discuss prior findings on how eye blinking patterns are associated with task difficulty and mental workload.

### 2.1 Eye Tracking and Interactive Systems

Eye tracking is the process of measuring eye movement, pupil size and blinking patterns. This has helped to facilitate our understanding of human mental processes over decades, e.g. how we process information during reading [59], the attentional process [2], and the neurobehavioral relationship in early human development [4]. With portable head-mounted setups, eye tracking has also supported a variety of human-computer interaction and usability studies [22,52,41,76,29]. This can be of use in identifying usability bottlenecks that increase completion time in web or app-menu designs [29,52]. Moreover, it provides physically disabled users with an alternative input modality [41] and helps design attentive user interfaces that track and analyze users' visual attention [76].

Advances in computer vision techniques have further enabled camera-based optical tracking to be widely adopted in many interactive systems [1,27,68,74,81], such as gaze-aided input for smart glasses [1] and virtual reality headsets [68]. In particular, facial landmarks detection performance (including eye regions) has been significantly improved given the large scale datasets collected from the wild [19,44,67]. This has helped to turn a simple RGB camera into a contact-free, low-cost but sophisticated eye tracker that can be used in many situations [27,74,81]. For example, Sugano *et al.* exploited this to explore audiences' collective gaze distributions on public displays and to measure their attention [74]. Zhang *et al.* investigated how this capability could be used to monitor eye contacts between persons [81]. In these fascinating interactive applications, eye movement and gaze patterns have significantly contributed to the provision of advanced user interaction features, with blinks still playing a minor role, e.g. supporting text entry [3].

### 2.2 Physiological Monitoring of Mental Demand

Physiological monitoring can help assess a user's mental states continuously, which is unique when compared with subjective self-reports and observational methods [16,27,36,50,66]. Given this, many researchers have turned to physiological computing, often with machine learning, to explore the relationship between physiological responses and mental demand. Of the many different types, cardiovascular, perspiratory and respiratory physiological signatures have been extensively explored in the literature [16,36,38,61].

Given the complexity of their relationship, multiple physiological sensors have often been exploited together to compensate for the limited sensitivity of each standard physiological metric (e.g. heart rate, breathing rate) [36,38,46]. However, the mentioned strategy may not be ideal for some real-world scenarios that require low sensing overhead. This has led recent research into advancing physiological metrics and understanding how to improve mapping between a unimodal physiological signature and mental demand [14,20,27,47], with breathing being a fitting example. Indeed, Cho *et al.* investigated how a representation of breathing can help read dynamic respiratory physiological responses to mental stressors for mental overload monitoring [14]. Eye tracking has also been investigated in light of this. Kosch *et al.* proposed to capture smooth pursuit eye movements to estimate cognitive load (working memory and visual demand) [47]. Further, Duchowski *et al.* developed a new metric to quantify pupillary responses to mental workload associated with task difficulty [20]. End-to-end learning was also investigated to automatically extract eye-movement patterns influenced by cognitive load tasks from eye images [27]. The present paper aims to contribute to this body of work by delving deeply into eye-blinking, which is the other remaining eye tracking measure.

**2.3  Spontaneous Blinking in relation to Mental Workload**

Spontaneous blinking is a physiological process that happens unconsciously, and differs from the two other types of blinking types, namely reflex (in response to environmental effects) and voluntary blinking (given internal effort) [11]. With spontaneous Blink Rate (BR, the number of blinks per unit time) and Blink Duration (BD, how long a blink lasts) as key standard metrics, spontaneous blinking has been studied in order to understand how it is related to information processing [26] and dopaminergic activities (regulating motor and limbic functions) [58]. For example, it has been reported that BR reduces during reading while it increases during speech compared with silence or relaxation [43].

Studies have explored the sensitivity of both standard metrics to participants' mental workload associated with different cognitive load factors in mentally demanding tasks (e.g. memory test) [6,13,25,40,60]. Despite slightly different definitions of the terms, mental workload and cognitive load factors (mainly, task difficulty and time pressure [28]), used across disciplines [28,35,47], it has been reported that BR is not sensitive to cognitive load aligned with increasing task difficulty itself, but that is sensitive to perceptual load (e.g. visual search and driving tasks with visual stimuli) [13,25,40]. In addition, a study reported that BD is more sensitive to visual demand than BR [6]. However, this metric also appears to be prone to interaction effect between task difficulty and visual stimuli [60]. The above suggests that the standard eye-blink metrics are not reliable in measuring task difficulty which is amongst the key components in usability assessment [50,62]. One of the possible reasons is that the hand-engineered metrics tend to oversimplify the complex physiological responses as reported with other physiological metrics (e.g. individual metric for heart rate variability [10,16]). This paper is concerned with how to capture dynamic patterns present in eye-blinking response and the extent to which task difficulty-associated mental workload is reflected on it.

**3   RETHINKING EYE-BLINK**

This section introduces *Rethinking Eye-blink*, a novel framework for capturing complex eye-blink patterns related to task difficulty over time and contributing to the continuous and automatic estimation of a user's task difficulty levels. Our fundamental module is a physiological representation of spontaneous blinking which aims to address the limited sensitivity of the standard eye-blink metrics (BR and BD) to task difficulty. To fully automate the process, the proposed framework includes three stages. First, it automatically extracts blinking timeseries from an RGB vision camera in real time. Second, it transforms the one-dimensional timeseries into a two-dimensional spectrogram that condenses dynamic patterns in blinking in the time-frequency domain. Third, it uses the represented pattern as an input to a multi-dimensional long-short term memory network for feature learning to estimate task difficulty levels. Although many of the technical components used in this framework have been previously proposed independently, our proposed framework uniquely empowers the eye-blink physiological signature to be better mapped onto mental workload parameters.

### 3.1 Automatic Extraction of Blink Timeseries

In our study, we use a built-in webcam as the main sensing channel to monitor eye-blinking patterns. Figure 2 illustrates how blink timeseries data can be extracted from said camera. Face landmark detection [44][1] is used to extract a person's eye shape and calculate the eye aspect ratio [72] (ratio of width to height of the eye, Figure 2a). An important note on handling the raw shape and ratio directly is that blinking does not accompany a simple binary reaction, but consists of onset and offset eyelid movements, as well as their duration (see Figure 2b). This can often be periodic, allowing frequency analysis that we describe in the next section. Given the potential ethical consideration, masking is applied to the other facial areas, leaving the outline of each eye. The ratio is affected by certain parameters, including angle, distance from camera, and head motion, thus leading to baseline wandering that often occurs in other physiological signals such as blood volume pulse [16]. To minimize this, we filter the raw signal (with 1 second moving average, empirically chosen) and subtract the filtered one from the raw signal (Figure 2c) as in [16], outputting blink timeseries.

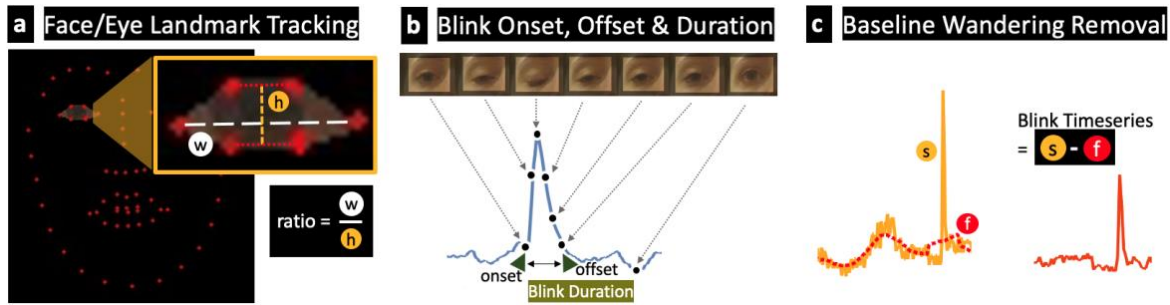

Figure 2: Automatic Blink Timeseries Extraction: (a) we use Face Landmark detection [44] to extract the shape of an eye from a standard webcam that captures images of a user's face continuously (masking is applied to non-eye regions). We then calculate the ratio of the width w to the height h as in [72] from the 6 identified landmarks for the eye in each frame, (b) the extracted ratio patterns along with eye shape changes in blinking (each peak accompanies onset and offset eyelid movements, as well as their duration) and (c) baseline wandering removal by subtracting the filtered signal from the extracted signal.

### 3.2 Time-Frequency Representation: Spectrogram Generator

The second stage focuses on a time-frequency representation of spontaneous blinking as summarized in Figure 3. This approach is inspired by the findings in our earlier study, where the Respiration Variability Spectrogram (RVS) was originally proposed [14]. For generating eye-blink spectrograms, we use a short time Gaussian window that slides across the extracted one-dimensional signal (Figure 3 left). Each time window signal is bandpass-filtered (BPF) to minimize noises and motion artefacts. Using Lomb-Scargle periodogram, which is powerful when it comes to handling irregularly sampled signals [75] (common in the real world), we calculate the Lomb-Scargle power of the filtered signal and stack it over time, resulting in a spectrogram (Figure 3). In this work, we set the blink range of interest to [0.033Hz, 0.4167Hz] (2 and 25 blinks per minute) heuristically, given the inconsistency in reported blink rate ranges [4,18]. The values are used as cutoff frequencies for the filtering and the size of each spectrogram. The first cutoff frequency determines the upper time limit of blinking (30 seconds), setting the window length to the upper limit multiplied by 2, followed by the addition of 1 (60+1 seconds). The Lomb-Scargle power of the signal in the sliding window is calculated every single second (1s step length). Figure 3 (bottom) shows an example of represented eye-blink patterns.

---

[1] For the implementation, we used the Dlib library and pre-trained model available at http://dlib.net/.

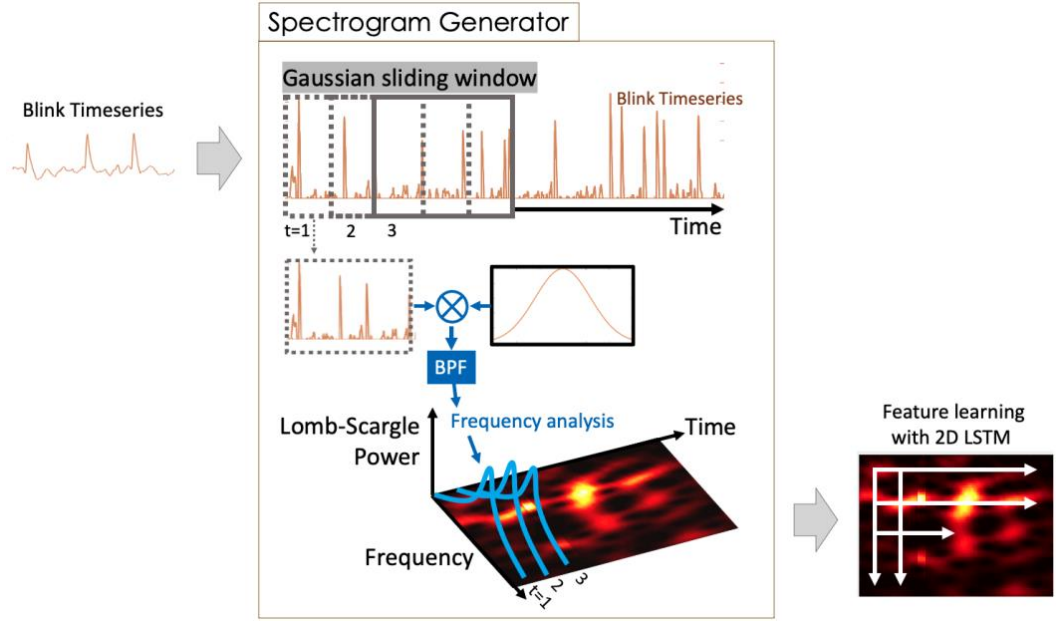

Figure 3: The spectrogram generator analyzes the extracted one-dimensional blink timeseries with a short time Gaussian sliding window. Each window filters the signal and produces frequency responses (Lomb-Scargle power). Each power spectrum is accumulated on the time-frequency plane to produce a spectrogram. This can then be fed into a machine learning model (e.g. 2D LSTM in this paper).

### 3.3 Feature Learning with 2D Long-Short Term Memory

The aim of this final stage is to approximate task difficulty levels from the proposed physiological representation of eye-blinks without the need for hand-crafting metrics from it. The use of convolutional neural networks (CNNs) has been proposed in many studies on end-to-end learning for a mapping from complicated physiological patterns to psychophysiological states [14,27,45,79]. Given that a CNN-based model has a large number of parameters in general, it requires a large dataset or extra data augmentation [14]. Otherwise, it is susceptible to overfitting, leading to poor performance on unseen data [73]. Although some techniques, such as dropout [73], could help address this issue, CNNs still have limited capability when it comes to learning sequential information and relationships. Given this, our work uses multi-dimensional long-short term memory (LSTM) for feature learning (Figure 1 right and Figure 3 right). LSTM is amongst the most powerful recurrent neural networks, and is designed to analyze sequential relation over a long period of time while helping to address vanishing gradients [37].

Here, a single 2D LSTM [31] is used for supervised learning for the mapping from spectrograms to task difficulty labels (both from subjective and objective measures). In our implementation for multi-class classification, the 2D LSTM layer consisted of 16 hidden units, which was linked to a fully connected layer. As in [31], the softmax activation was used at the layer with the cross-entropy loss function. The 2D input (each represented blink pattern) to the network was normalized given our interest in the dynamic (but possibly hidden) patterns of blinking present within each measurement block. This was done through feature scaling with min-max normalization. The above process enhances patterns on each spectrogram and minimizes varying ranges of frequency amplitudes across spectrograms in a similar way as a model benefits from batch normalization dominantly used in CNNs [39]. This feature learning model was implemented using Tensorflow 2.0[2]. The implementation of the *Rethinking Eye-blink* framework is available at *https://www.youngjuncho.com/2021/rethinking-eyeblink*.

## 4 STUDY I: SENSITIVITY OF BLINK METRICS TO TASK DIFFICULTY

The aim of our first study is to revisit standard eye-blink metrics and examine their sensitivity to task difficulty parameters in a lab experiment, as well as to understand the extent to which a representation of blinking improves the sensitivity. We used a mental arithmetic task with different difficulty levels. To control other factors that affect blinking patterns, we fixated a central target (to minimize visual stimulus and perceptual load [13,40]) with a fixed time duration for each question (to control time pressure which is another cognitive load factor [28]), while we also controlled ambient lighting conditions [56].

### 4.1 Experimental Design

We adopted the Mathematical Serial Subtraction test [71], as this is one of the standard protocols in exploring *task difficulty* effects across fields [20,69]. As described in Figure 4, this protocol is divided into two main task blocks with different difficulty levels: i) easy level (a fixed single-digit number as a subtractor, e.g. 1), ii) hard level (a varying two-digit prime number, e.g. 13, 17). The two blocks, which are counterbalanced (within-group design), are presented after a relaxation baseline block. During both the task blocks, each subtraction question is present in the center of the screen for 7.5 seconds. In the baseline, participants are asked to relax while seeing the center of a black screen. Note that having the three condition blocks, specifically the independent variable conditions in this study, is also linked to multi-class classification tasks which we explore in Study 2. The study protocol is approved by the University College London Interaction Centre ethics committee (ID Number: UCLIC/1617/003).

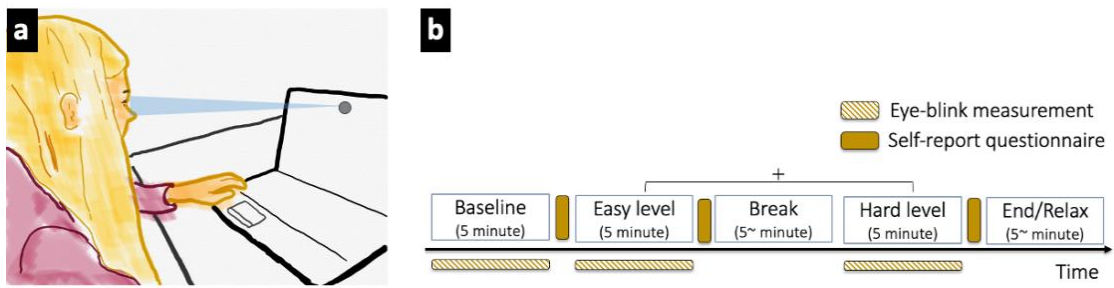

Figure 4: (a) Experimental setup: only the standard webcam embedded in a laptop is used to monitor participants' eye-blink patterns and apply the *Rethinking Eye-blink* framework to the data collection, (b) experiment workflow mainly consisting of baseline, easy and hard level blocks (+ denotes counterbalancing).

---

[2] https://www.tensorflow.org/

### 4.2 Participants and Materials

18 participants (4 females and 12 males) participated in this study. 4 out of the 18 participants (22.2%) wore glasses. Participants were recruited through the university participant pool mailing lists and local communities. Prior to data acquisition, participants were given both an information sheet and a consent form. Moreover, participants were informed that they could leave the study at any given point. This study was conducted in a quiet laboratory room which contained a comfortable chair and a neat desk; there was no window in the room, the purpose of which was to facilitate the controlling of ambient lighting. An experimenter was present in the room but did not observe participants' performance, so as to avoid introducing social evaluative threats given our focus on task difficulty assessment rather than on stress [16].

The tasks were implemented in MATLAB. Due to interpersonal differences in working memory capacity that influence participants' perception of task difficulty [20,77], the difficulty levels for the second task were designed to adapt to each participant's performance. We programmed this by choosing a subtractor from a set of prime numbers adaptively along with a participant's latest correct answer rates (e.g. the last three questions); for instance, once a user answered correctly for three times in a row, the subtractor was set to a higher prime number, e.g. from 13 to 17.

During both tasks and the baseline period, a built-in webcam (720p FaceTime HD camera) was used to record each participant's face, and the *Rethinking Eye-blink* framework (Figure 4a) was employed to measure blinking patterns in a non-contact manner. Here, we downsampled the recorded frames (640x480) to lower the computational cost. Each block lasted for 300 seconds. Given participants' possible interaction with the experimenter at the beginning and the end of each block (e.g. head turned), we used 260 seconds of data in the middle. During the blocks, participants' actual performances on the math tests (proportion of correct responses) were automatically recorded. After each block, participants were given a questionnaire to measure their perceived task difficulty levels and asked to take a rest as much as they wanted, so that they could feel relaxed, as well as to minimize eye fatigue (minimum 5 minutes). The questionnaire used a 10-cm visual analog scale (VAS), which allowed participants to rate on a continuous scale, so as to avoid nonparametric properties in the data [16,49]. Instead of presenting fewer relevant workload components used in standard questionnaires such as physical demand in NASA TLX [35], we mainly asked, *"How difficult was the task?" (ranging from 0, not at all, to 10, very much)* to directly capture *perceived* task difficulty levels.

As main dependent variables in this experiment, the two-standard eye-blink metrics were used: BR (Blink Rate) and BD (Blink Duration). Given the complexity in quantifying dynamic physiological responses [10], a single metric hardly quantifies our proposed blink spectrogram properly. However, designing a numerical metric for this is required in order to statistically compare its sensitivity with the standard metrics. Hence, in this study we propose to use Shannon's entropy [65,78] of a represented blink spectrogram to examine how random the pattern is. The purpose of this is to ascertain if there are certain patterns inside each spectrogram for a different task (e.g. a low entropy value indicates that it does have patterns [78]). We name this metric as Blink Entropy (BE) which can be expressed as:

$$BE(X) \equiv -\sum_{(i,j)} p(x_{i,j}) \log_2 p(x_{i,j}) \tag{1}$$

where $x_{i,j}$ is the amplitude of element (time=$i$, frequency=$j$) on a spectrogram and $p(x_{i,j})$ is the probability distribution estimated with a normalized histogram [78].

### 4.3 Results

Figure 5a shows the distributions of the self-reported perceived task difficulty across each block (Baseline: M=0.34, SD=0.37; Easy: M=1.77, SD=1.23; Hard: M=6.15, SD=1.47). We analyzed the subjective ratings to ascertain whether the protocol was valid in inducing considerable differences of task difficulty for each block. As the data from the baseline and easy task were skewed (p=0.009, p=0.011 from the Shapiro-Wilk test), we conducted the Friedman test. The results showed a significant effect of the task type on the perceived task difficulty levels ($\chi^2(2)$=34.11, p<0.001). The Wilcoxon signed-rank test confirmed that there were significant differences across blocks (Baseline-Easy: Z=-3.552, p<0.001; Baseline-Hard: Z=-3.729, p<0.001; Easy-Hard: Z=-3.725, p<0.001), thus indicating that the experimental protocol was designed properly to investigate the relationship between task difficulty and blink measures.

The distributions of blink metrics are summarized in Figure 5b (left: standard metric, BR; middle: standard metric, BD; right: new metric, BE). Overall, it is clear that the new blink metric (BE) for our proposed time-frequency representation is much more sensitive to task difficulty than the standard metrics. Some examples of the blink spectrogram samples from the three blocks are shown in Figure 6. The blink spectrograms from the baseline showed high entropy (randomness) while the other two presented certain patterns with low entropy (also see Figure 5b right). For significance tests, we carried out a one-way repeated measures ANOVA on BE, and applied the Friedman test to both BR and BD (given that some sets of the data were not normally distributed). While there were no significant effects of task difficulty on either of the standard metrics (BR: $\chi^2(2)$=1.778, p=0.411; BD: $\chi^2(2)$=2.333, p=0.311), there was a significant difference in BE across tasks (F(2,34)=6.203, p<0.01). The post-hoc paired t-test with Bonferroni correction showed that BE had significant differences between Baseline and Easy (p=0.04), as well as between Baseline and Hard (p=0.031). This confirmed the stronger sensitivity of the entropy-based metric from our proposed representation, despite no significant difference between Easy and Hard, indicating that the pattern strength levels (from the randomness measure) are still not informative in discriminating between both blocks (events).

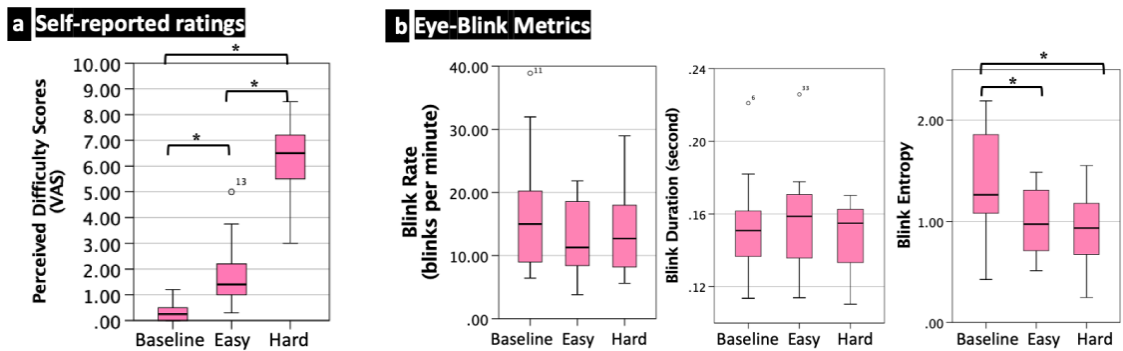

Figure 5: Boxplots (95% confidence interval) of (a) perceived difficulty scores from the VAS-based questionnaire and (b) eye-blink metrics (Blink Rate, Blink Duration and Blink Entropy) of 18 participants across blocks (Baseline, Easy and Hard task blocks). Asterisk marks (*) denote significant differences (p<0.05). ° denotes outliers.

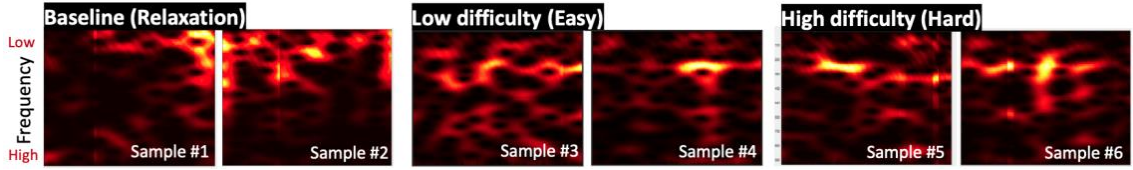

Figure 6: Examples of blink spectrograms from each block (left: Baseline; middle: Easy task block; right: Hard task blocks).

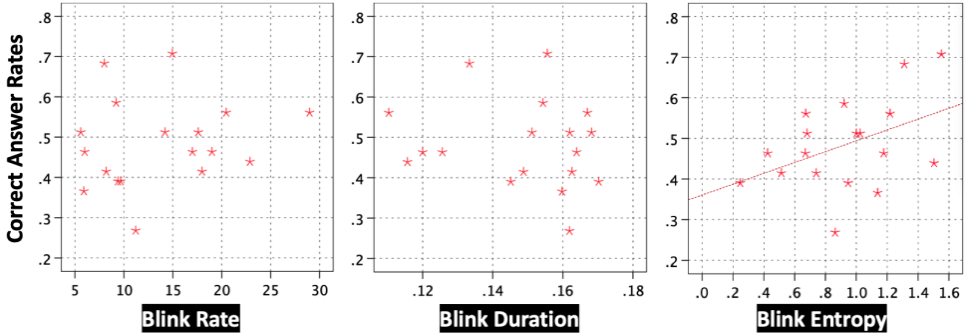

Figure 7: Scatter plots of the correct answer rates (objective task difficulty measure) and the three eye-blink metrics (left: Blink Rate; middle: Blink Duration; right: Blink Entropy with a dotted linear regression line).

Furthermore, we investigated the capability of each metric in determining different difficulty levels along with subjective (perceived difficulty scores) and objective measures (correct answer rates) separately. First, we tested the correlations of the perceived task difficulty with each metric. The perceived scores were collected from all three blocks. Given some non-normally distributed data, we used the non-parametric Spearman's correlation coefficient. BE maintained a much higher correlation with the self-report than did BR and BD (BE: r=-0.341, p=0.012; BR: r=-0.017, p=0.903; BD: r=-0.011, p=0.935). To compensate for the subjectivity and inter-personal difference in the perceived measure, we also tested the correlations of the investigated metrics with participants' correct answer rates (proportion of correct responses to math questions) as an objective measure. As the Easy block produced significantly higher correct answer rates and lower difficulty perception, we focused solely on the data only from the Hard block. As shown in the scatter plots in Figure 7, BE maintained a stronger correlation with the correct rate (with approaching significance) than other metrics (BE: r=0.437, p=0.07; BR: r=0.176, p=0.485; BD: r=-0.169, p=0.504), thus confirming its higher sensitivity to task difficulty.

## 5 STUDY II: TASK DIFFICULTY ESTIMATION

Study I has confirmed the importance of exploring time-frequency representation of eye-blink when capturing its response to task difficulty. In this section, we aim to evaluate the capability of our proposed *Rethinking Eye-blink* framework in estimating task difficulty levels. As described in Section 3, a 2D LSTM is used in the framework for a non-linear mapping from blink spectrograms to difficulty-related parameters (Figure 1).

### 5.1 Dataset, Labeling Strategy and Validation

We used the dataset from Study I. The dimension of each spectrogram was 200 (time unit, note: a 61-second sliding window converting each 260-second blink timeseries to a spectrogram of 200 seconds) x 93 (frequency elements in the range [0.033Hz, 0.4167Hz]). Labeling each instance from every participant is an important step in machine learning classification. In this work, we adopted three labelling strategies: i) event-based, ii) subjective measure (perceived difficulty score)-based, and iii) objective measure (correct answer rate)-based labeling. The event-based labeling involved using three task blocks (Baseline, Easy and Hard) directly as class labels. This has often been exploited in the literature (e.g. [47]) for multi-class classification.

In the subjective and objective measure-based labeling strategies, we focused only on the arithmetic task blocks to examine the capability of our framework in differentiating between low and high difficulty levels (binary classification). The self-reported scores were used to create the two sets of classes for the subjective labeling, as in [16,38,61]. This is interesting because *intended* difficulty levels (e.g. easy and hard tasks) are sometimes mismatched with perceived levels (e.g. P13 in Figure 5a). Likewise, the correct answer rates were used to create two sets of classes for the objective difficulty labeling. Following our earlier work [14,16], the K-means clustering was used to split each measure into two classes (k=2) separately given that it helps to better handle interpersonal variability in labeling data when compared to simply using a middle value for separation [16].

To evaluate the person-independent classification performance of our method with the three labeling strategies, we trained and tested it using an 18-fold leave-one-subject (person or participant)-out (LOSO) cross-validation. The LOSO cross-validation has been widely used to test the ability to generalize to unseen participants' physiological patterns [16,47]. At each fold, blink spectrograms from the participants, except for one participant, were used to train the machine learning model, and the unseen data from the left-out participant was employed for testing. Given its purpose for the generalization, this type of cross validation tends to produce lower inference performance than simple splitting (e.g. 80% for training and 20% for testing) and k-fold cross-validation (e.g. mean multilabel classification accuracies reported in [47]: 48.9% from LOSO, 73.3% from non-LOSO).

For fair comparisons, we also implemented other baseline methods using the hand-engineered blink metrics (i.e. BR, BD, BE) explored in Study I. The baseline methods were as follows: Support Vector Machine (SVM) with linear kernel, Random Forest (RF), and k-Nearest Neighbor (kNN), all of which are widely used in this field. Note that we employed grid search for hyperparameter optimization to ensure the best performance from these baseline methods. Furthermore, we included ablation studies by taking the one-dimensional eye-blink timeseries as low-level features (instead of 2D eye-blink spectrograms in Figure 3) as in [16]. They were input into Multi-Layer Perceptron (MLP) and 1D LSTM separately to analyze the importance of each proposed component. Given our well-balanced dataset, we used mean accuracy over the 18 folds as our evaluation metric. Each method was run on 2 GPUs (GeForce RTX 2080 Ti) with a 64-bit Windows machine (Intel Core i9-7900X 3.3GHz 128GB RAM).

### 5.2 Results

Table 1 shows mean accuracies from the 18-fold LOSO cross-validation of each implemented method along with the three labeling strategies. Overall, the proposed method (eye-blink spectrogram representation + 2D LSTM) achieved over 70% mean accuracies for all cases, outperforming the other baseline methods. In detail, in the event-based multi-class classification, where the majority case (the ratio of the number of instances in the majority class to the total number) was 33.3%, our method produced a mean accuracy of 70.4% - much higher than the chance level for the three classes. This was also 18.5 percentage points higher than the best performance from the baseline method (SVM with linear kernel and hand-engineered features, mean accuracy = 51.9%).

From the binary classification tasks, our method produced a mean accuracy of 72.2% for the subjective measure (self-reported perceived difficulty)-based labeling, and a mean accuracy of 77.8% for the objective difficulty case (correct answer rates). Both scores showed an improvement of 8.3% and 11.1% over the best baselines for subjective and objective labeling strategies, respectively (best baseline methods: SVM + hand-engineered features for *subjective*; 1D LSTM + Eye-Blink timeseries for *objective*). Interestingly, the implemented methods tended to perform better under the objective labeling than the subjective one.

From the comparisons of our method with the other LSTM approach, where we instead used the eye-blink timeseries as input, we observed that our proposed eye-blink representation generated much higher performance for all cases. This result confirms that the sensitivity gain is indeed from the more powerful representation. Overall, the *Rethinking Eye-blink* framework demonstrated strong performance in estimating task difficulty levels from different perspectives, comparable with recently-reported results from other types of mental demand inference tasks with the LOSO cross-validation [16,47].

Table 1: Mean accuracies over the 18-fold LOSO (Leave-One-Subject-Out, also called Leave-One-Person-Out) testing sets from each method along with three labeling strategies.

| Input | Method | Labelling Strategy | | |
|---|---|---|---|---|
| | | Event-based (Multi-class) | Subjective Difficulty (Binary) | Objective Difficulty (Binary) |
| Hand-engineered features (BR, BD, BE) | SVM | 51.9% | 63.9% | 61.1% |
| | RF | 42.6% | 55.6% | 63.8% |
| | KNN | 46.3% | 61.1% | 58.3% |
| Eye-Blink timeseries | MLP | 40.7% | 50.0% | 52.8% |
| | LSTM | 48.2% | 58.3% | 66.7% |
| Proposed Eye-Blink Spectrogram | LSTM | **70.4%** | **72.2%** | **77.8%** |

## 6 APPLICATIONS

Before discussing the lessons learnt from the studies, we discuss some examples of the possible HCI applications of *Rethinking Eye-blink*.

### 6.1 Tools for Accessibility Assessment

The *Rethinking Eye-blink* framework can support automated and continuous assessment of mental workload associated with task difficulty. This can help understand hidden barriers for many user groups to the use of newly introduced features in computing systems and user interfaces. For example, this can be of use when it comes to identifying and correcting contributing factors to inaccessibility for certain user groups in navigating websites and mobile applications. Under the EU Directive 2016/2102[3], organizations and individuals in many countries are obligated to make web and mobile applications accessible; however, achieving this for every prospective user is a very challenging task. With existing tools for accessibility and usability evaluation, such as a single user's eye-gaze plot, and collective heatmaps [54] together, our proposed method can visually highlight what causes difficulties in navigation and performing certain tasks so as to enhance corresponding features and improve accessibility (Figure 8a).

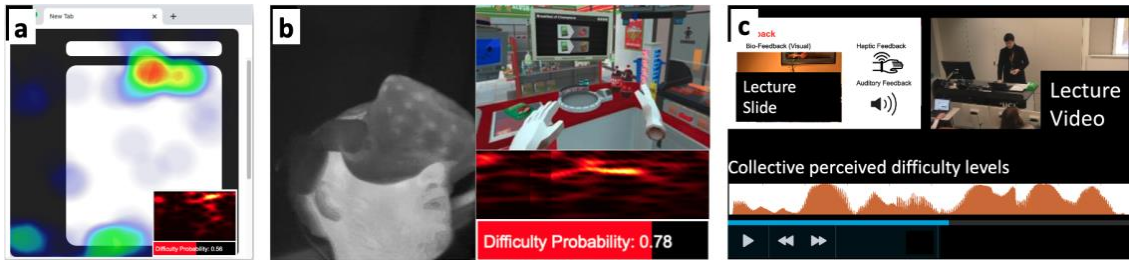

Figure 8: Example applications: (a) the Rethinking Eye-Blink framework can support automated and continuous evaluation of web and mobile accessibility with existing tools, such as a single user's eye-gaze plot and collective heatmaps [54] together, (b) this can also help understand hidden difficulty and accessibility issues with newly-introduced features in many fascinating VR applications, (c) perceived content difficulty levels can be estimated, collectively gathered, and visualized for tutors to tailor their teaching to students' mental demand during synchronized live-streamed teaching.

We used a built-in webcam as the main sensing channel in this study given its cost-effectiveness in regard to technology diffusion for supporting many contexts, including low resource settings [5]. Nonetheless, we also highlight that our approach can be sensing-interface-independent. This means that it can work with many wearable applications that use eye trackers, with Virtual Reality (VR) head-mounted display serving as an example. In VR, users interact with virtual objects to complete tasks and experience immersive virtual environments in a number of applications [12,70]. VR applications have been actively explored in HCI research for supporting creative processes [53], training for surgical trainees [33,48] and 360-degree cinematic storytelling [57]. With the market welcoming a growing number of VR headsets which embed an eye tracking system (e.g. HTC VIVE Pro Eye), we envision the use of this approach as shown in Figure 8b to understand hidden accessibility issues in such fascinating applications, in turn making them more engaging and inclusive.

### 6.2 Collective Online Classroom

Steeply increasing attention has been paid to online teaching and distance learning globally during the COVID-19 pandemic. A known issue in online learning environments is a learner's limited interaction with tutors and peers [51]. This often makes it difficult for tutors to perceive how difficult students may find the content, and to adjust their teaching phase and materials. Moreover, this demotivates learners when it comes to participating. Here, the ability to continuously probe learners' perceived difficulty levels may provide potential benefits to address this. For instance, as illustrated in Figure 8c, tutors may review students' collective experiences of difficulty (visualized with a line chart) in real time during synchronized live-streamed teaching. This can help tutors to understand which components challenge learners and to tailor teaching activities to students' mental demands. Learners may feel more empathy and encouraged to engage. It would be interesting to investigate the extent to which this type of intervention can facilitate the collective learning process.

---

[3] EU Directive (2016/2102) on the accessibility of the websites and mobile applications of public sector bodies (https://eur-lex.europa.eu/eli/dir/2016/2102/oj)

## 7 DISCUSSION

Understanding users' mental states is an important key to innovation in human computer interaction. In this paper, we have investigated spontaneous eye-blink responses to task difficulty and how a user's mental workload associated with it can be approximated automatically. To address the limited capability of standard blink metrics [58,60], we have proposed the *Rethinking Eye-blink* framework, which focuses on a representation of dynamic eye-blink patterns and feature learning to enhance the role of blink measures in task difficulty assessment. Our approach uses a standard webcam, without requiring any other additional sensors and interfaces that can be occasionally cumbersome or heavy to use in certain scenarios. The proposed framework can be easily integrated into many interactive systems in real-world contexts (e.g. home and indoor office environments) to examine usability and accessibility problems in a longitudinal way.

Results from our studies have demonstrated that the time-frequency representation of blinking can lead to a stronger association with the intended (i.e. event-based) and perceived (i.e. self-reported) task difficulty levels and users' performance (i.e. correct answer rates) than the standard metrics. This highlights the importance of capturing complicated physiological patterns from many different angles (e.g. as in heart rate variability analysis [16,64] and respiration variability [14]) to enhance their sensitivity to mental states-related factors (in our work, task difficulty). This is aligned with the arguments regarding the need for more advanced methods to compensate for certain standard physiological metrics that tend to oversimplify physiological phenomena [10,16,20]. Despite the wealth of research on eye-blink, there were also doubts about the usefulness of blinking in other tasks where simple metrics were primarily used (e.g. blink rate in [21,58]). Investigations into diverse patterns of blinking could possibly have led to different viewpoints in earlier studies.

Inferencing mental states through physiological measurements is a challenging task in general, given that a certain physiological state could be suggestive of different types of mental states [80]. Hence, many recent works in HCI have turned to machine learning for a better mapping of physiological signatures to psychological states [9,47,63,80]. Likewise, the *Rethinking Eye-blink* framework does include a machine learning classifier (2D LSTM) to benefit from its automatic feature learning capability. While the entropy metric that we designed to quantify the proposed blink spectrogram showed still limited capacity in differentiating between two different difficulty levels in Study I, the overall *Rethinking Eye-blink* framework led to clear improvements in performance and sensitivity. Furthermore, we highlight that person-independent classification with LOSO cross-validation (as in [16,47]) was investigated to test the generalizability of the aforementioned framework. With this, we envisage some benefit from the above types of machine learning approaches, whose pre-trained models in lab settings are scalable in real-life situations where datasets grow.

It is noteworthy that this approach is differentiated from other end-to-end learning approaches where raw image data is directly fed into a model. In most cases, the latter rely on a much larger dataset, often shortening the length of the input data to increase the instances. For example, 6-second eye image sequences were used as input in [27], which is not sufficiently long to reflect spontaneous blinking (e.g. average interval between blinks for men in a cognitive process is 20 seconds [7]); hence, many instances tend to exclude "blink" moments.

### 7.1 Limitations and Future Directions

Despite the promising results, we have identified opportunities for improving and extending this approach. Firstly, while this work focuses on spontaneous blinking in a lab setting, two other blink types (reflex and voluntary [11]) may interfere with the performance in daily situations. A possible solution would be to enable the proposed framework to adapt to each context and/or to be personalized so as to learn new rules from the data. However, in a technological intervention where a user is encouraged to blink voluntarily (e.g. mitigate eye fatigue [18]), the compatibility of our method with this may still be limited. It would be interesting to explore how such spontaneous blinking can be recognized against the other blink types automatically.

Secondly, our work did not take into account the other two eye tracking data categories – eye movement (e.g. saccades, fixations) and pupillary response – which were accessible in our dataset. This was because of our aim to rethink the role and usefulness of eye-blink measures in mental workload assessment and to investigate how to empower such a unimodal approach for better inference capability. Nonetheless, it would be a fascinating topic to investigate how multimodal eye tracking data can amplify mental status monitoring capability all together as in [9] (personality detection). Furthermore, other physiological signatures (e.g. blood volume pulse [16]) which can be measured from the single sensing channel (i.e. camera) can be incorporated into this type of system to explore a variety of human computer interaction scenarios in more depth, which the standard task used in this paper does not represent.

Lastly, although our approach does mask none-eye areas in the early stage of the processing, there are still privacy concerns regarding the use of an RGB-vision-based approach. One possible idea to address this is to use other imaging sensors that read electromagnetic radiation outside the visible range, such as far-infrared imaging sensors [15,17] which can possibly gauge a person's blinking but which are less identifiable. Machine learning algorithms may also contribute to this, abstracting other personally identifiable information.

## 8 CONCLUSION

This paper described *Rethinking Eye-blink,* a new framework that focuses on a physiological representation of eye-blink responses with a spectrogram generator to monitor a user's task difficulty in daily tasks. Our approach addresses the limited capacity of standard blink metrics in assessing mental workload bound with task difficulty. Connecting the generator to a recurrent neural network makes it possible to automatically learn features that are informative about task difficulty-related parameters, no longer requiring hand-engineering. The results from our first study have confirmed that the standard blink metrics (blink rate and blink duration) are not sensitive to task difficulty and we can significantly benefit from the representation of spontaneous blinking when capturing its responses to task difficulty. In the following classification study, our method yielded notably higher
performance in discriminating between different task difficulty levels (e.g. 77.8% mean accuracy from LOSO to test the generalization capability) than other baseline methods using the standard metrics. This underlines the feasibility of the proposed approach in the potential use scenarios discussed in the present paper. The implemented software tools are available for download at https://www.youngjuncho.com/2021/rethinking-eyeblink.

### ACKNOWLEDGMENTS

We thank the anonymous reviewers for their feedback. This work was supported in part by the UK Department for International Development through the AT2030 Programme (www.AT2030.org).

19